\documentclass[pra,twocolumn,showpacs,preprintnumbers,amsmath,amssymb]{revtex4-1}

\usepackage{graphicx}% Include figure files
\usepackage{dcolumn}% Align table columns on decimal point
\usepackage{bm}% bold math

\begin{document}

\title{Collective dynamics of vortices in trapped Bose-Einstein condensates}
\author{Tapio Simula}
\affiliation{School of Physics, Monash University, Clayton, Victoria 3800, Australia}

\begin{abstract}
We have calculated collective mode spectra for three-dimensional, rotating Bose--Einstein condensates in oblate harmonic traps using the microscopic Bogoliubov--deGennes field theory. For condensates with $N_v$ vortices, $N_v$ Kelvin--Tkachenko mode branches are obtained. The features of these modes are compared with those predicted by a classical point vortex model. We have created movies to visualize the motion of the vortices corresponding to the Kelvin--Tkachenko waves. 
\end{abstract}

\maketitle

% Intro ================================================================================
\section{Introduction}
Vortical flows stand out among the most fascinating realizations of hydrodynamics. In equilibrium, an isolated vortex filament often settles into the shape of a straight line. Perturbations to such equilibrium cause helical vortex displacement waves to propagate along the length of the vortex filament. These excited states of vortices are known as Kelvin waves, named after Lord Kelvin who derived a dispersion relation for such vortex excitations \cite{Thomson1880a}. Pitaevskii predicted similar vortex excitation modes to exist in superfluids \cite{Pitaevskii1961a}. Kelvin waves of quantized vortices have been observed both in superfluid helium \cite{Ashton1979a} and in trapped atomic Bose--Einstein condensates \cite{Bretin2003a}. However, measurement of the long-wavelength superfluid kelvon dispersion relation remains an experimental challenge.

For two or more vortex filaments, the stability of different vortex configurations becomes a matter of importance. Although rarely realized in classical hydrodynamical systems, the quantization of circulation in quantum systems, together with energetic considerations, leads to the emergence of stable vortex lattice configurations in quantum liquids. Considering equilibrium states of flux vortices in superconductors, Abrikosov predicted them to be arranged in a triangular lattice structure \cite{Abrikosov1957a}, corresponding to the most efficient way of packing cylinders. In addition to superconductors, triangular Abrikosov vortex lattices also emerge in rotating scalar superfluids \cite{Donnelly1991a,Anderson2010a}. For vector superfluids more exotic lattice structures appear \cite{Fetter1983a,Salomaa1987a,Parts1995a,Mizushima2004a,Huhtamaki2009a,Simula2011aa,Simula2012a}.

The stability of multivortex configurations against perturbations was addressed in the classical context by Thomson who studied the stability of 3 through 7 vortices arranged in a single orbital ring pattern \cite{Thomson1883a}. Havelock extended this result finding the 7 vortex case to be marginally stable having an excitation mode with precisely zero frequency (modes with negative frequencies corresponding to instabilities), further showing that an arbitrary number of vortices in a ring configuration can be stabilized by adding sufficiently strong vortex in the centre of the ring \cite{Havelock1931a}. An exhaustive analysis of the transverse normal modes of finite vortex arrays and their stability was presented by Campbell \cite{Campbell1981a}. Tkachenko studied the stability of an infinite triangular vortex lattice in a superfluid helium and predicted the existence of a transverse normal mode oscillation now known as Tkachenko wave \cite{Tkachenko1966a}. Due to the end-cap effects in rotating superfluid helium systems, the modes calculated by Campbell \cite{Campbell1981a} did not facilitate a quantitative comparison with the photographic movies of few-vortex excitation modes observed by Yarmchuck and Packard \cite{Yarmchuck1981a}.
 
The experimental realization of dilute-gas Bose--Einstein condensates \cite{Anderson1995a,Davis1995a,Bradley1995a} opened up the possibility to controllably create and image individual quantum vortices \cite{Matthews1999b} and vortex lattices \cite{Madison2000a,Abo-Shaeer2001a,Hodby2001a,Haljan2001a} and to study the normal mode oscillations of such vortex arrays \cite{Anderson2010a}. The vortex precession observed in the experiment by Anderson \emph{et al.} may be viewed as fundamental Kelvin wave motion of a vortex in a Bose-Einstein condensate \cite{Anderson2000a}. Bretin \emph{et al.} excited a Kelvin mode of higher axial wave number by exploiting its resonant coupling to the quadrupole mode of the condensate \cite{Bretin2003a}. Tkachenko waves were created by Coddington \emph{et al.} in rapidly rotating vortex lattices \cite{Coddington2003a}. Smith \emph{et al.} observed a gyroscopic tilting mode of a vortex array \cite{Smith2004a} which has one axial node and belongs to the common mode Kelvin--Tkachenko mode branch \cite{Simula2010aa}. Recently, Weiler \emph{et al.} \cite{Weiler2008a} observed dynamics of vortex dipoles and Freilich \emph{et al.} applied continuous imaging methods to obtain trajectories of such vortex-antivortex dipoles \cite{Freilich2010a}. In three-dimensions such vortex dipoles may break into vortex loops via the Crow instability \cite{Spreiter1951a,Crow1970a,Berloff2001a,Simula2011a}. Similar techniques could be used to experimentally discover the complete set of low-lying collective Kelvin--Tkachenko mode branches of few-vortex arrays.

Here we calculate the low-lying normal modes of oblate, three-dimensional, Bose--Einstein condensates hosting up to 19 vortices, using the microscopic Bogoliubov--de Gennes theory. This model fully accounts for the internal structure of the vortex cores and the coupling of vortex motion to the compressional sound waves. We solve the equations governing the three-dimensional excitation modes of such vortex arrays without forcing any symmetries in the calculations. The Kelvin--Tkachenko mixed modes of few-vortex single orbital arrays for three-dimensional prolate condensates were calculated in \cite{Simula2010aa}. Here we extend those results to oblate systems with vortex lattices having up to 3 concentric vortex orbitals.

% FIGURE ===
\begin{figure}
\includegraphics[width=\columnwidth]{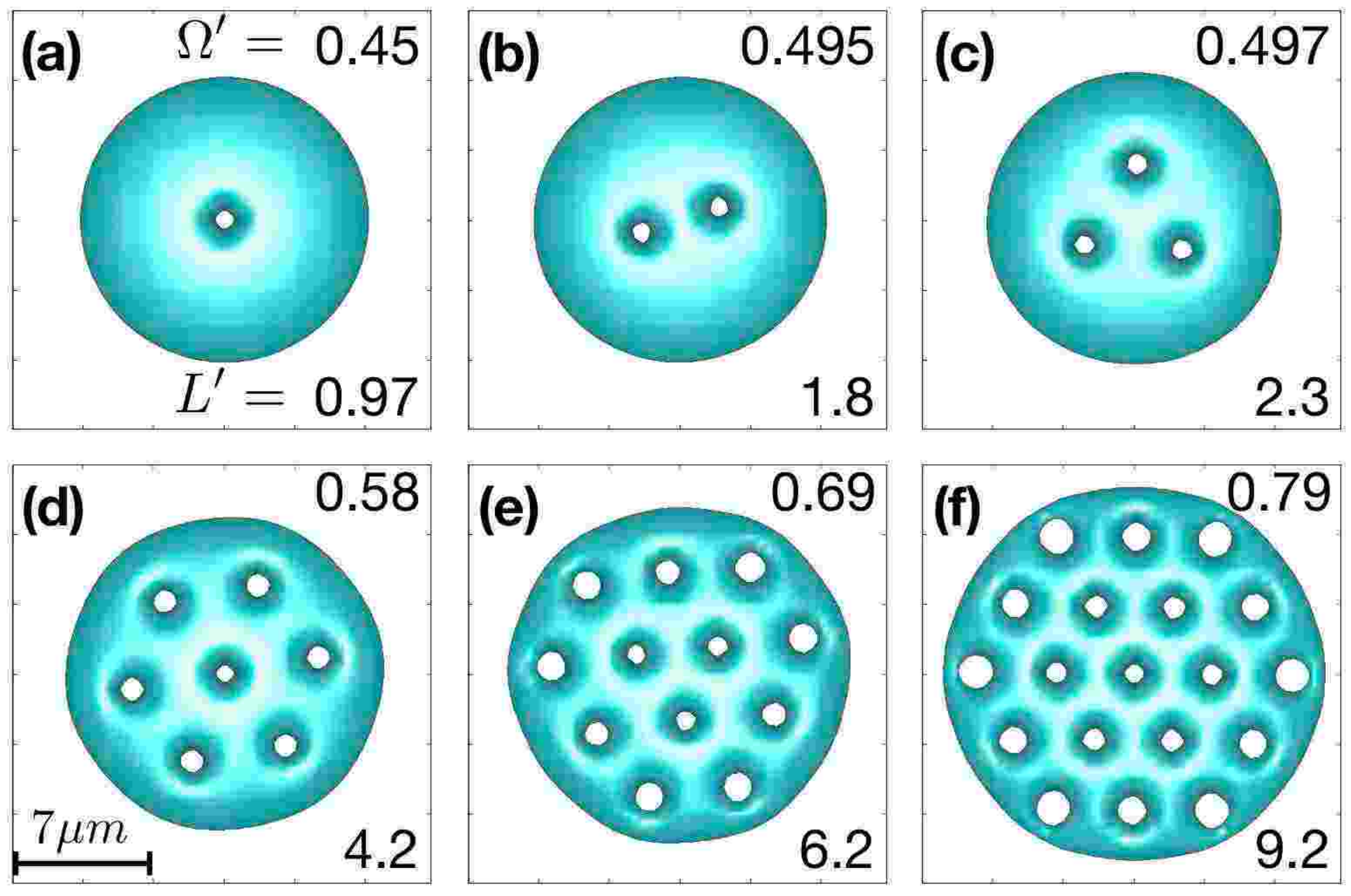}
\caption{(Color online) Ground states of a Bose-Einstein condensate in rotating harmonic traps. Each plot is a density isosurface viewed along the rotation axis. The states (a)-(f), respectively accommodating 1, 2, 3, 7, 12, and 19 vortices, are being rotated at respective angular frequencies, $\Omega' =\Omega/\omega_\perp =$ 0.45, 0.495, 0.497, 0.58, 0.69 and 0.79. The expectation values of the orbital angular momentum per particle $L'=\langle L\rangle /(\hbar N)$ of these ground states are 0.97, 1.8, 2.3, 4.2, 6.2, and 9.2 respectively.} 
\label{fig1}
\end{figure}
% FIGURE ===

% Model ================================================================================
\section{Model}
We employ the Bogoliubov--deGennes (BdG) wavefunction formalism to calculate the excited states of the three-dimensional harmonically trapped rotating Bose--Einstein condensates whose multiply connected ground state wavefunction $\phi({\bf r})$ contains $N_v$ single quantum vortices. This amounts to solving the eigenvalue problem
\begin{equation}
\begin{pmatrix}
\mathcal{L}({\bf r})-\mu & g\phi^2({\bf r})\\
-g\phi^*({\bf r})^2 & -\mathcal{L^*}({\bf r})+\mu
\end{pmatrix}
\begin{pmatrix}
u_q({\bf r}) \\
v_q({\bf r})\end{pmatrix}
=\hbar\omega_q
\begin{pmatrix}
u_q({\bf r}) \\
v_q({\bf r})
\end{pmatrix},
\label{bdg}
\end{equation}
where $u_q({\bf r})$ and $v_q({\bf r})$ are the complex valued quasiparticle wavefunctions labelled by $q$, $\omega_q$ are the corresponding eigenfrequencies and the chemical potential $\mu=\langle \mathcal{L}({\bf r}) -g|\phi({\bf r})|^2\rangle$ where $\mathcal{L}({\bf r})$ is a single particle operator defined in Eq.~(\ref{GPL}) and $g$ is a coupling constant. The numerical challenges encoded in Eq.~(\ref{bdg}) are largely due to the near-singular spectrum of the BdG operator and the fact that for three-dimensional systems the matrix representation of Eq.~(\ref{bdg}) becomes very large. For single vortex lines, cylindrical symmetry is often deployed to reduce the computational cost of solving Eq.~(\ref{bdg}) \cite{Simula2001d,Virtanen2001a}. However, when more than one vortex is present in the system, the cylindrical symmetry of the ground state is lost and one has to solve the full three-dimensional problem, which allows correct treatment of the axial structure of the vortex modes. We have used a finite-element discrete variable representation \cite{Schneider2006a,Simula2008d} to obtain a sparse matrix representation for the BdG operator of this three-dimensional system and have used a parallelized Arnoldi iteration \cite{PARPACK,Simula2010aa} to solve the eigenvalue problem.

The dynamics of the calculated collective modes is visualized by tracking the evolution of the perturbed condensate density $ |\phi_q({\bf r},t)|^2$  as a function of time $t$, where \cite{PitaevskiiStringari2003a}
\begin{equation}
\phi_q({\bf r},t) = \left[ \phi({\bf r}) +  p \delta{\phi}_q({\bf r},t)\right] e^{-i\mu t/\hbar} 
\label{vis1}
\end{equation}
and the perturbation caused by the excitation of the collective mode $q$ is
\begin{equation}
\delta{\phi}_q({\bf r},t) =  u_q({\bf r})e^{-i\omega_q t} + v^*_q({\bf r}) e^{i\omega_q t}.
\label{vis2}
\end{equation}
The population of the quasiparticle excitation mode is controlled by the number $p$ in Eq.~(\ref{vis1}). We have used $p=0.3$ in this paper.

% Model: GP ================================================================================
Prior to solving Eq.~(\ref{bdg}) we must obtain the rotating ground state wavefunction $\phi({\bf r})$ by 
solving to high accuracy the Gross-Pitaevskii equation
\begin{equation}
\left(\mathcal{L}({\bf r})-g|\phi({\bf r})|^2 -\mu\right) \phi({\bf r}) = 0
\label{GP}
\end{equation}
where the operator
\begin{equation}
\mathcal{L}({\bf r}) = -\frac{\hbar^2\nabla^2}{2m}+V_{\rm ext}({\bf r})+2g|\phi({\bf r})|^2  - \Omega L_z
\label{GPL}
\end{equation}
and the constant $g=4\pi\hbar^2 a/m$ determines the strength of $s-$wave interactions between particles of mass $m$ in terms of the scattering length $a$. The particles are confined by an external harmonic potential 
\begin{equation}
V_{\rm ext}({\bf r})=m \left(\omega^2_\perp r^2 + \omega_z^2z^2\right) /2
\end{equation}
with transverse $\omega_\perp$ and axial $\omega_z$ frequencies. The system is rotated at an angular frequency $\Omega$, and the $z$-component of the orbital angular momentum operator is $L_z= -i\hbar(x\partial_y-y\partial_x) $. The normalization of the wavefunction, $\int_V|\phi({\bf r})|^2 d{\textbf r}=N$, determines the number of particles $N$ in the system. Here we consider $N \approx 3\times10^4$ $^{87}$Rb atoms confined in a harmonic trap with an axial trap frequency $\omega_z=2\pi\times 100$Hz and an aspect ratio $\omega_z/\omega_\perp = \sqrt{8}$. The dimensionless interaction parameter $gN=1000\;\hbar\omega_\perp a_0^3$, where $a_0=(\hbar/m\omega_\perp)^{1/2}$.

\section{Kelvin-Tkachenko modes}
To facilitate the discussion of the collective vortex displacement modes, the corresponding ground states $\phi({\bf r})$ need to first be  specified. Figure \ref{fig1} shows condensate density isosurfaces viewed along the rotation axis for six different vortex systems, calculated using Eq.~(\ref{GP}). The reduced external rotation frequency $\Omega' = \Omega/\omega_\perp$ and the expectation value of the reduced orbital angular momentum $L' = \langle L_z\rangle/(N\hbar)$ are marked in the upper and lower right corners of each frame, respectively. The rotation causes the condensate to expand radially due to the centrifugal effect. Simultaneously the cloud becomes flatter in the axial direction. This centrifugal effect is fairly small in the single orbital cases (a)-(c) but becomes prominent for the two (d) and (e) and three orbital (f) vortex arrays. Although the superfluid circulation of the vortices is quantized in integer multiples of $h/m$, the orbital angular momentum per particle $\langle L_z\rangle/N$ is not an integer multiple of $\hbar$ and hence the quantization of orbital angular momentum is not exact despite the topological discreteness of the vortices. In contrast, for axisymmetric multiquantum vortices the orbital angular momentum is quantized in integer units \cite{Simula2002a}.

% FIGURE ===
\begin{figure}[!t]
\includegraphics[width=\columnwidth]{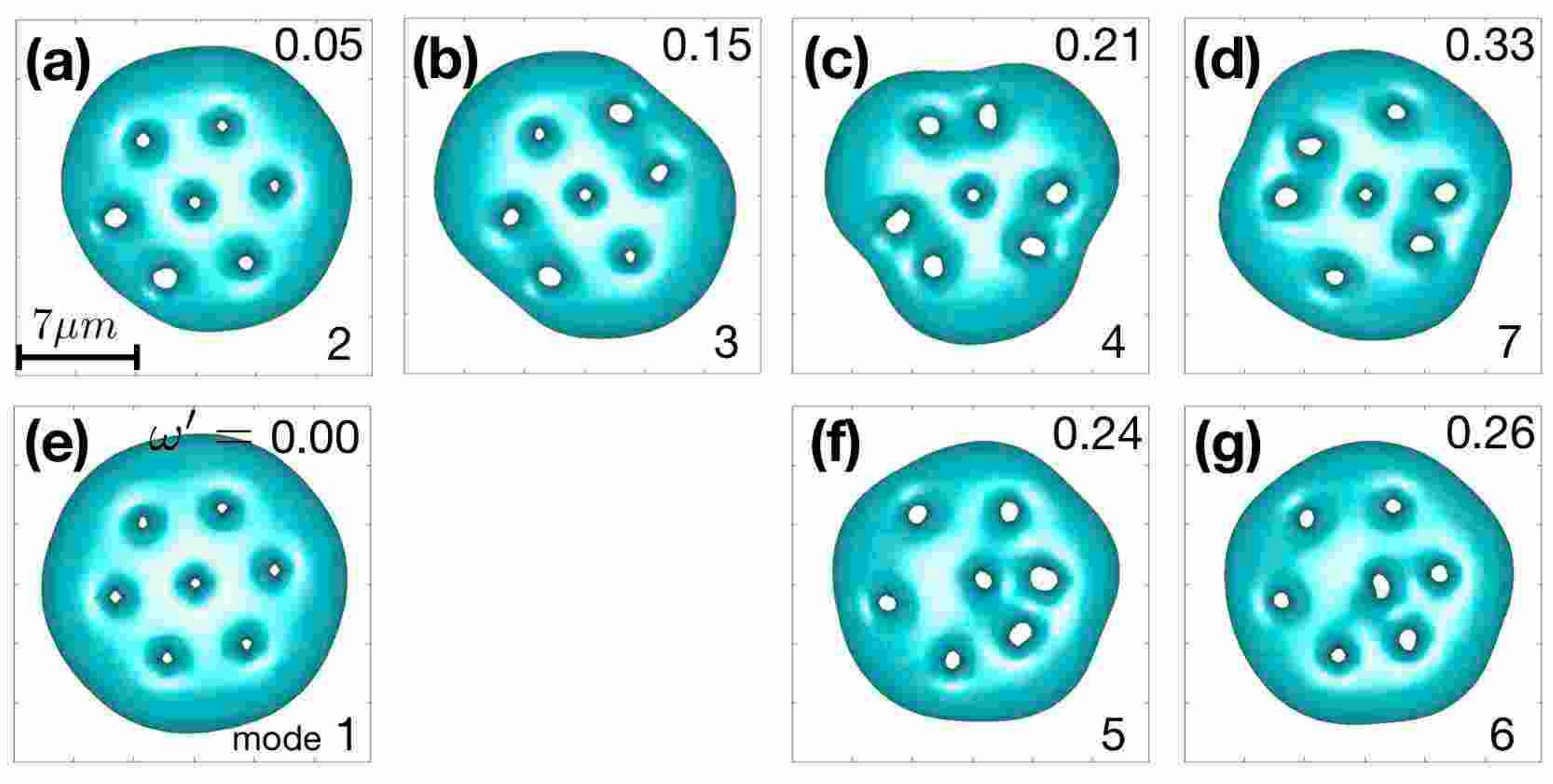}
\caption{(Color online) Instantaneous density isosurfaces of the condensate with 7 vortices, perturbed by the different lowest vortex displacement modes. The view is along the rotation axis. The excitation frequency $\omega'$, expressed in units of the transverse trap frequency, of each perturbing mode is indicated in the upper right hand corner of each frame. The discrete rotation symmetries of these modes are inherited from the sixfold symmetry of the triangular equilibrium vortex lattice. The central vortices in (f) and (g) gyrate in the opposite directions [see supplementary material \cite{movies}]. The number in the bottom of each frame corresponds to the mode label in Table I.} 
\label{fig4}
\end{figure}
% FIGURE ===

\subsection{General considerations}
Let us first restrict our discussion to the subset of modes $(N_v,n_z=0)$, which have no axial nodes. For such modes all vortices remain straight and parallel to each other and the vortex dynamics is effectively two-dimensional, although the excitation frequencies do depend on the full three dimensional structure of the condensate. Irrespective of the number of vortices $N_v>0$ in the ground state, two vortex modes, which we refer to as (C) and (T), always appear in these systems. 

The simplest vortex displacement mode shared by all vortex configurations is the common mode or center-of-mass excitation (C) of the vortices, in which the whole vortex array executes retrograde circular motion, in the rotating frame of reference, around the trap center as a rigid body. Here we have deviated slightly from the terminology of Campbell \cite{Campbell1981a}. By the common mode we refer to the fact that the vortex motion is common to all vortices in a single mode, where as Campbell uses the term common mode to refer to fact that similar vortex mode exists for all vortex configurations irrespective of the number of vortices in the system. Apart from the obvious difference in excitation energy, the (C) mode may appear similar to the usual counter-rotating dipole mode, in which the vortex lattice stays stationary and the condensate density executes circular motion around the trap center. In contrast, in the vortex common mode (C), which only exists for $N_v\ne0$, the lattice rotates with respect to the trap and the stationary condensate.  

The second universal collective excitation is the transverse Tkachenko mode (T), which typically has the lowest excitation energy in the system and reveals itself as an azimuthal torsional oscillation of the vortex lattice \cite{Andereck1980a,Rajagopal1964a,Tkachenko1966a,Fetter1975a,Williams1976a,Sonin1976a,Sonin1987a,Baym1983a,Baym2003a,Gifford2004a,Coddington2003a,Simula2004a,Mizushima2004b,Woo2004a,Baksmaty2004a,Simula2010aa}. In this mode, the individual vortices travel (in the rotating frame of reference) along elliptical periodic orbits around their own equilibrium positions, their combined motion resulting in the collective mode of the whole vortex lattice. 

In addition to these two universal vortex displacement modes, for systems with at least two concentric vortex orbitals Campbell \cite{Campbell1981a} finds ``quadratic" (Q) modes in which the central single vortex or cluster of vortices executes rigid body oscillation while the vortices in the outermost orbital are able to move relative to each other. There are two different quadratic (Q) modes corresponding to the opposite sense of rotation of the central vortex cluster. Further following the notation of Campbell \cite{Campbell1981a} we label the modes, for which the central vortex remain stationary, as rational (R) modes.

We observe that a generic collective vortex mode can be constructed by first grouping the vortices into sub-arrays and then considering the motion of the vortices in each sub-array as an independent collective mode. The vortex motion of the whole lattice is then constructed as a superposition of the collective modes of the sub-arrays. This is further illustrated in the end of the Sec. III.

The underlying ground state vortex lattice has 6-fold symmetry, which is reflected by the $n-$fold symmetry of the Bogoliubov modes \cite{Mizushima2004b}. Furthermore, each Kelvin--Tkachenko branch can be uniquely labelled using a set of $N_v$ vortex displacement phase shifts $\Theta_i$ for the vortices \cite{Simula2010aa}. However this, like any means of uniquely identifying the individual modes, quickly becomes cumbersome for large numbers of vortices. 

We also note that the orbital motion of a single vortex around the trap centre admits three equivalent descriptions. It can equally be viewed as the long wavelength axially nodeless $n_z=0$ Kelvin wave, the single vortex Tkachenko wave, or the rigid body common mode of the vortex lattice with only one vortex.

% FIGURE ===
\begin{figure}[!t]
\includegraphics[width=\columnwidth]{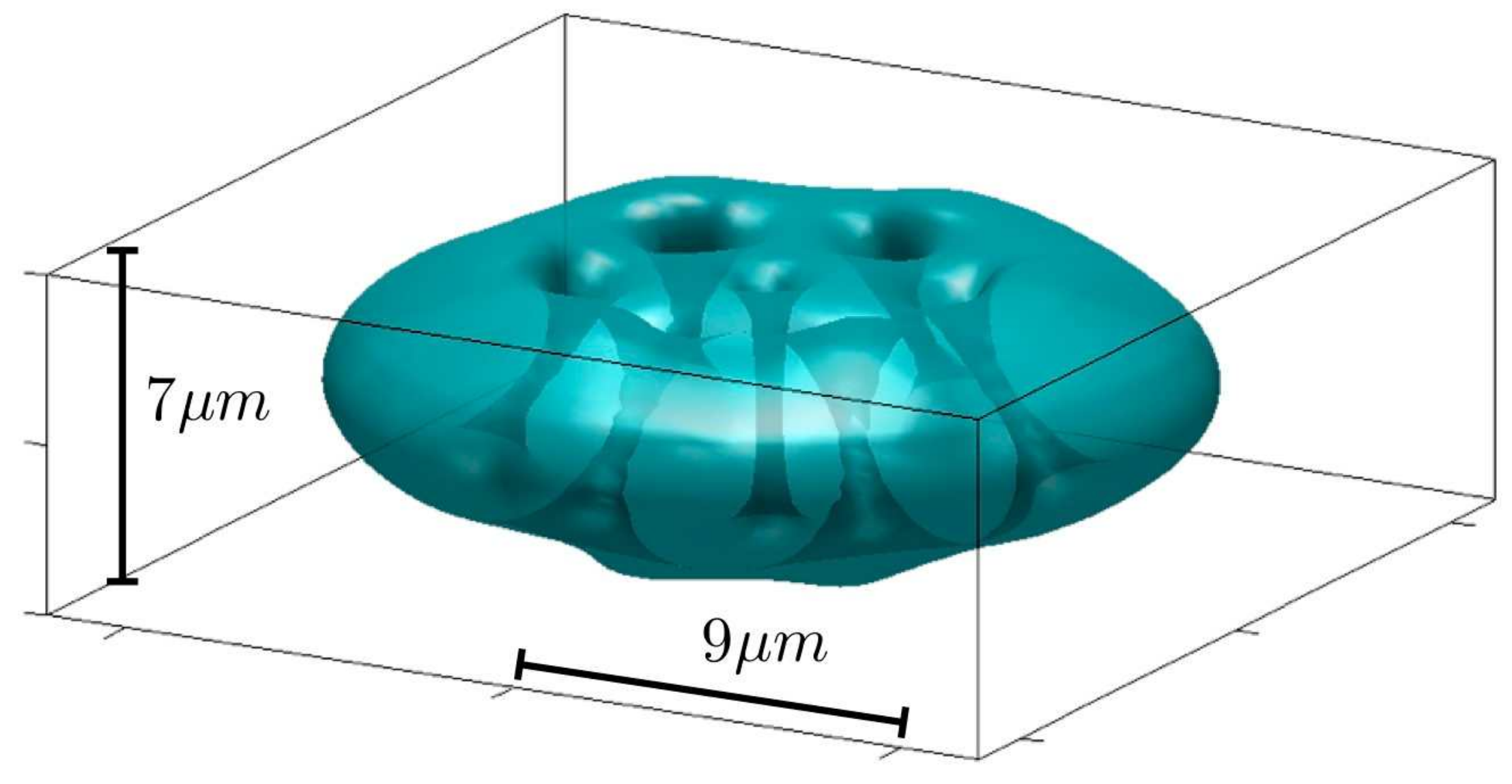}
\caption{(Color online) Instantaneous density isosurface of a condensate excited in a first axial $n_z=1$ Kelvin--Tkachenko mode corresponding to the mode branch 1 of the seven vortex system in Table I. The orbital vortices are symmetrically tilted toward the central vortex. For the full dynamics of this mode we refer to the attached movies \cite{movies}. } 
\label{fig5}
\end{figure}
% FIGURE ===

\subsection{Details of vortex modes}
Table I lists the frequencies of the calculated axially nodeless $n_z=0$ vortex displacement modes for vortex numbers $N_v= 1, 2, 3, 7, 12,$ and $19$. The Kelvin--Tkachenko mode branches are sorted according to increasing energy of the lowest mode in each branch and are referred to by integers. The pure Tkachenko modes (T), common modes (C), quadratic modes (Q) and rational modes (R) are labeled in the table. We have produced a movie featuring the collective dynamics of the vortices for each tabulated mode, see the supplementary material \cite{movies}.

The motion of $N_v$ vortices in two-dimensions can be approximated using a classical point vortex model \cite{Helmholtz1858a,Onsager1949a}
\begin{equation}
i\Gamma_\nu\frac{\partial z_\nu}{\partial t} = \frac{\partial H}{\partial z_\nu},
\label{pv}
\end{equation}
where $z_\nu=x_\nu+iy_\nu$ are complex functions of the spatial vortex coordinates $x_\nu$ and $y_\nu$, $\Gamma_\nu$ are the strengths of the vortices, and $H=-\sum_{\nu\ne\eta}\Gamma_\nu\Gamma_\eta \log(|z_\nu-z_\eta|)/4\pi$ is the potential driving the vortex motion when there are no boundaries. For confined systems more general form of $H$ is used. This is a set of $N_v$ first order differential equations. Seeking small perturbations to the vortex positions leads to $N_v$ by $N_v$ matrix equation with $N_v$ eigenvalues corresponding to the $N_v$ normal modes of the vortex system. Splitting Eq.~(\ref{pv}) into two real valued equations 
\begin{eqnarray}
\Gamma_\nu\dot{x}_\nu &=& \partial H /\partial{y_\nu} \notag\\
\Gamma_\nu\dot{y}_\nu &=& -\partial H/\partial{x_\nu},
\label{pv2}
\end{eqnarray}
yields a system with $2N_v$ eigenmodes, only half of which are linearly independent the other half being related to the former by a simple rotation \cite{Campbell1981a}. From Eqns~(\ref{pv2}) it can be seen, by associating $x_\nu$ and $y_\nu$ with the canonical coordinate and momentum of a Hamiltonian system, that the phase space of this 2D vortex system corresponds to that of an equal number of 1D massive particles. For a Newtonian point mass particle, the resultant force acting on it is proportional to the acceleration of the particle and therefore in two-dimensions a system of $N_p$ particles have $2N_p$ normal modes. However, here the Magnus force, which drives the motion of a vortex, is proportional to the velocity and therefore the vortex system has only $N_v$ independent low-lying normal modes. 

Equation.~(\ref{pv2}) can also be used to reproduce the vortex motion of the axially nodeless Kelvin-Tkachenko collective modes by appropriately choosing the vortex potential and the initial vortex positions. However, the resulting point vortex dynamics of this chaotic Hamiltonian dynamical system is highly sensitive to the initial vortex locations. It therefore appears that these few-vortex superfluids may be particularly useful systems for studies of quantum chaology \cite{Berry1987a} and, in particular, for the study of the role of wavefunction phase singularities to the behavior of quantum systems whose classical counterparts exhibit chaos. The point vortex model was recently used to obtain quantitative agreement with experimental observations for the case of a single vortex-antivortex pair in a harmonically trapped Bose-Einstein condensate \cite{Middelkamp2011a,Torres2011a}.

\squeezetable
\begin{table}[!t]
\caption{Axially nodeless $n_z=0$ excitation frequencies $\omega_q$ of vortex configurations with $N_v$ vortices. Tkachenko (T), common mode (C), quadratic (Q) and rational (R) modes are marked by the respective labels. See supplementary movies for the dynamics of each tabulated mode \cite{movies}. }
\begin{center}
\begin{tabular}{ccccccccccc}
\hline\hline
$N_v$ & mode & $\omega_q (\omega_\perp)$ & label  &&&&$N_v$ & mode & $\omega_q (\omega_\perp)$ & label\\
\hline
1 & 1 &  0.223 & T,C    &&&&     &   9 &  0.284 &  \\
   &    &            &          &&&&      & 10 & 0.329 &  \\
2 & 1 &  0.157 &  T      &&&&      & 11 & 0.394 &  \\
   & 2 &  0.202 &  C      &&&&     & 12 & 0.420 &   \\
   &    &            &          &&&&      &      &            &   \\
3 & 1 &  0.068 &  T      &&&& 19 & 1   & 0.020 & C \\
   & 2 &  0.143 &          &&&&      & 2   & 0.022 & T,R \\ 
   & 3 &  0.280 &  C     &&&&      & 3   & 0.079 &  R  \\
   &    &            &          &&&&      & 4   & 0.139 & R\\
7 & 1 &  0.000 &  T,R   &&&&      & 5  & 0.157& Q\\
   & 2 &  0.048 &  C     &&&&      & 6   & 0.200 & Q\\
   & 3 &  0.146 &  R      &&&&      & 7   & 0.210 & R\\
   & 4 &  0.213 &  R      &&&&      & 8   & 0.258 & R\\
   & 5 &  0.236 &  Q      &&&&      & 9  & 0.260 & R\\ 
   & 6 &  0.263 &  Q     &&&&      & 10 & 0.265 & \\
   & 7 &  0.329 &  R     &&&&      & 11 & 0.300 & R\\
   &    &            &          &&&&      & 12& 0.307& \\ 
12& 1 & 0.009 &  T      &&&&     & 13 & 0.360 & \\ 
   & 2 &  0.026 &  C     &&&&      & 14 & 0.363 & R\\
   & 3 &  0.101 &          &&&&      & 15 & 0.380 & R\\
   & 4 &  0.150 &          &&&&      & 16 &  0.394& \\
   & 5 &  0.180 &  Q     &&&&      & 17 &  0.418& \\
   & 6 &  0.222 &  Q     &&&&      & 18 & 0.498 & \\
   & 7 &  0.266 &          &&&&      & 19 & 0.501 & \\
   & 8 &  0.268 &          &&&&      &      &   & \\
   \hline
   \end{tabular}
\end{center}
\label{default}
\end{table}%

\subsubsection{Single orbital modes}
The Kelvin--Tkachenko modes of 1, 2, and 3 vortex single orbital systems are qualitatively equivalent with those obtained for elongated $\omega_z/\omega_\perp = 0.2$ condensates \cite{Simula2010aa}. Briefly, the single vortex system has only one Kelvin--Tkachenko mode branch corresponding to the Kelvin waves propagating along the axis of the quantum vortex \cite{Svidzinsky2000a,Simula2008a,Koens2012a,Koens2012b}. The varicose vortex waves \cite{Simula2008b} do not involve the motion of the vortex phase singularities and are not discussed further here. The two-vortex system has two mode branches corresponding to the in-phase (C) and out-of-phase (T) motions of the two vortex filaments. In addition to the Tkachenko and common modes, the three-vortex system has a mode branch with $2\pi/3$ phase shift between each pairing of vortices \cite{Yarmchuck1981a,Simula2010aa}. Both the axially nodeless $n_z=0$ and first axially excited states $n_z=1$ of the single orbital vortex systems are included in the supplementary videos \cite{movies}. 

Using the notation ($N_v$, mode branch, $\omega_q/\omega_\perp$) the first axially excited $n_z=1$ Kelvin--Tkachenko modes have frequencies (1,1,1.532), (2,1,1.536), (2,2,1.658), (3,1,1.600), (3,2,1.690), and (3,3,1.605). The frequencies of these modes are approximately half of the harmonic oscillator level spacing in the axial direction, $\omega_z/2=\sqrt{2}\omega_\perp$. By tightening the trap sufficiently in the axial direction, practically all axial Kelvin--Tkachenko modes can be suppressed \cite{Rooney2011a}. Dynamics of the modes for the four-vortex system were presented in \cite{Simula2010aa}.

% FIGURE ===
\begin{figure}
\includegraphics[width=\columnwidth]{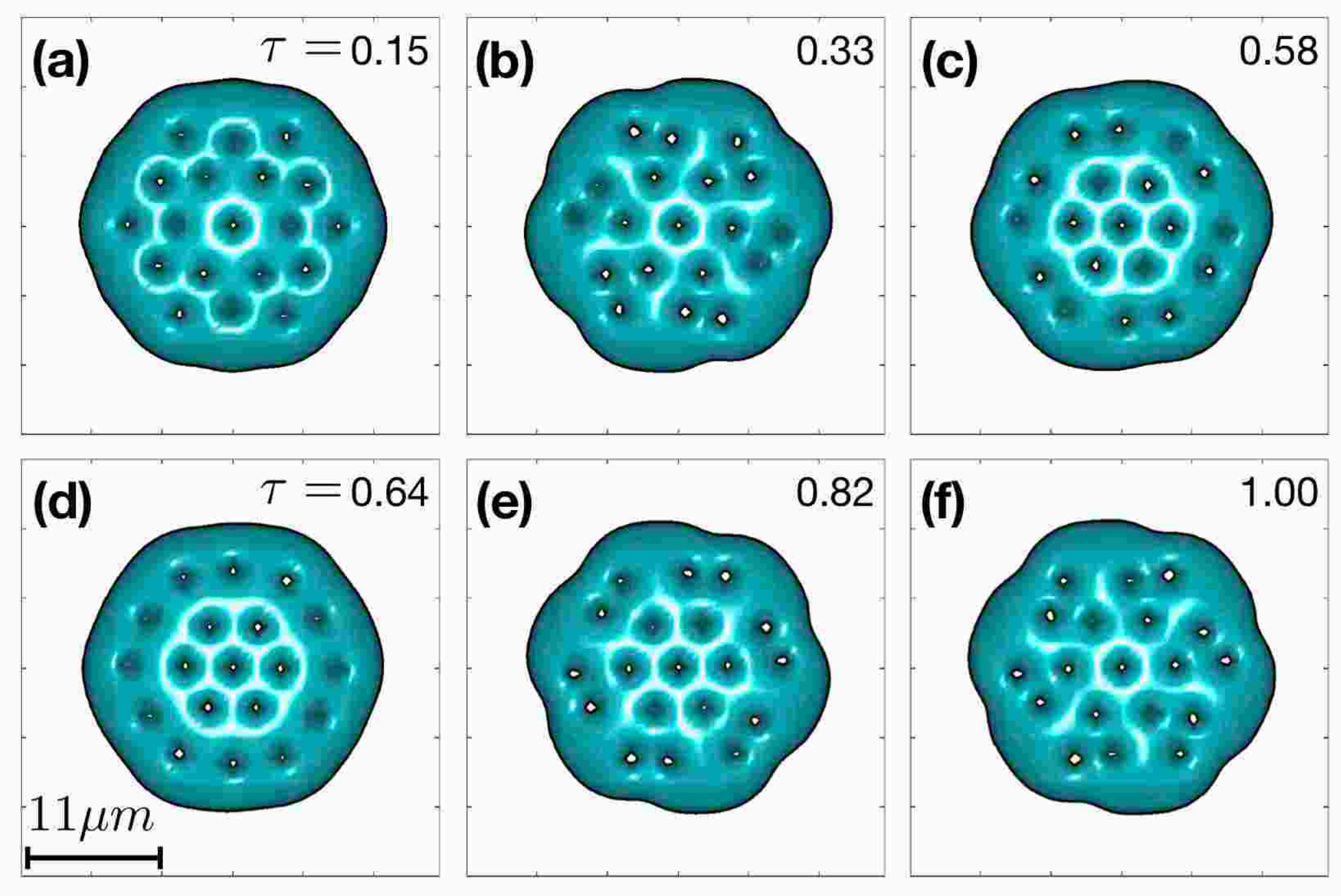}
\caption{(Color online) Snapshots of the time evolution of the condensate with 19 vortices and perturbed by the rational mode 14 in Table I. The times $\tau$ marked in the frames (a) - (f) are in units of $2\pi/\omega_q$.} 
\label{fig6}
\end{figure}
% FIGURE ===

\subsubsection{Multiorbital modes}
The 7-vortex system is special. Out of all stable two orbital vortex arrays it has the smallest number of vortices. It is also known that 7 classical vortices distributed along a single ring pattern is the marginally stable case such that any higher number of vortices will be unconditionally unstable \cite{Thomson1883a,Havelock1931a}. Figure \ref{fig4} (a)-(g) shows snapshots of the condensate perturbed by each of the $N_v=7$ different axially nodeless $n_z=0$ Kelvin--Tkachenko modes. The excitation frequency of these modes is marked in the upper right corner of each frame. Frame (a) is the common (C) mode, (e) is the Tkachenko (T) mode, (f) and (g) are the two quadratic modes (Q) and (b)-(d) are the two rational (R) modes, see Table I. 

\begin{figure}
\includegraphics[width=1\columnwidth]{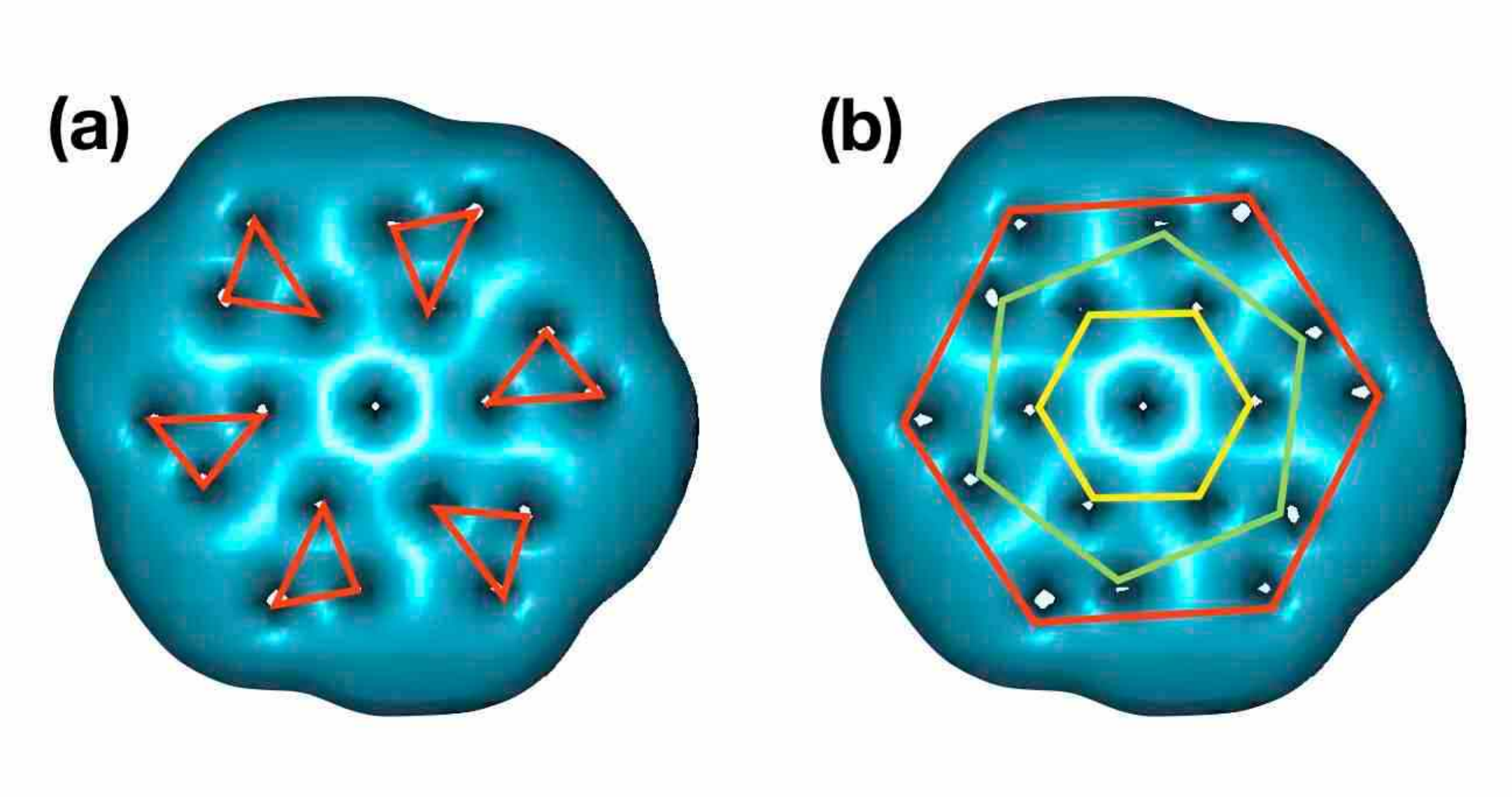}
\caption{(Color online) Two different decompositions of the same collective mode of a 19 vortex array into collective modes of sub-arrays. (a) six three vortex modes (b) three seven vortex modes.} 
\label{fig7}
\end{figure}

Figure \ref{fig5} shows a density isosurface of a condensate perturbed by the lowest axially excited Kelvin--Tkachenko mode corresponding to the same branch as the mode in the frame (e) (cf. $N_v=7$, mode 1 in Table I). Here the central vortex remains straight while the other vortices are tilted performing gyroscopic motion about their respective equilibrium positions. For the characterization of the collective vortex modes with 12 and 19 vortices we refer the reader to the movies included in the supplementary material \cite{movies}. Figure \ref{fig6} illustrates one such mode for the case of 19-vortex, three-orbital, configuration ($N_v=19$, mode 14 in Table I). Figures \ref{fig6} (a) -(f) show snapshots of the condensate density isosurface perturbed by this mode for times $\tau=$ \{0.15, 0.33, 0.58, 0.64, 0.82 and 1.00\}$\times 2\pi/\omega_q$. The vortex motion in this mode can be broken down to six equivalent three-vortex clusters, see Fig.~\ref{fig7}(a), each executing a three vortex collective motion (Table I, $N_v=3$, mode 2) while the central vortex remains stationary. Alternatively, it can be viewed as a superposition of three Tkachenko modes (Table I, $N_v=7$, mode 1), the 12 vortices in the outermost orbital being divided into two interlaced groups and the inner 6 vortex orbital forming the third sub-array and the central vortex being shared by all three sub-arrays, see Fig.~\ref{fig7}(b). 

% FIGURE ===
% FIGURE ===

% FIGURE ===
\begin{figure*}
\includegraphics[width=1.7\columnwidth]{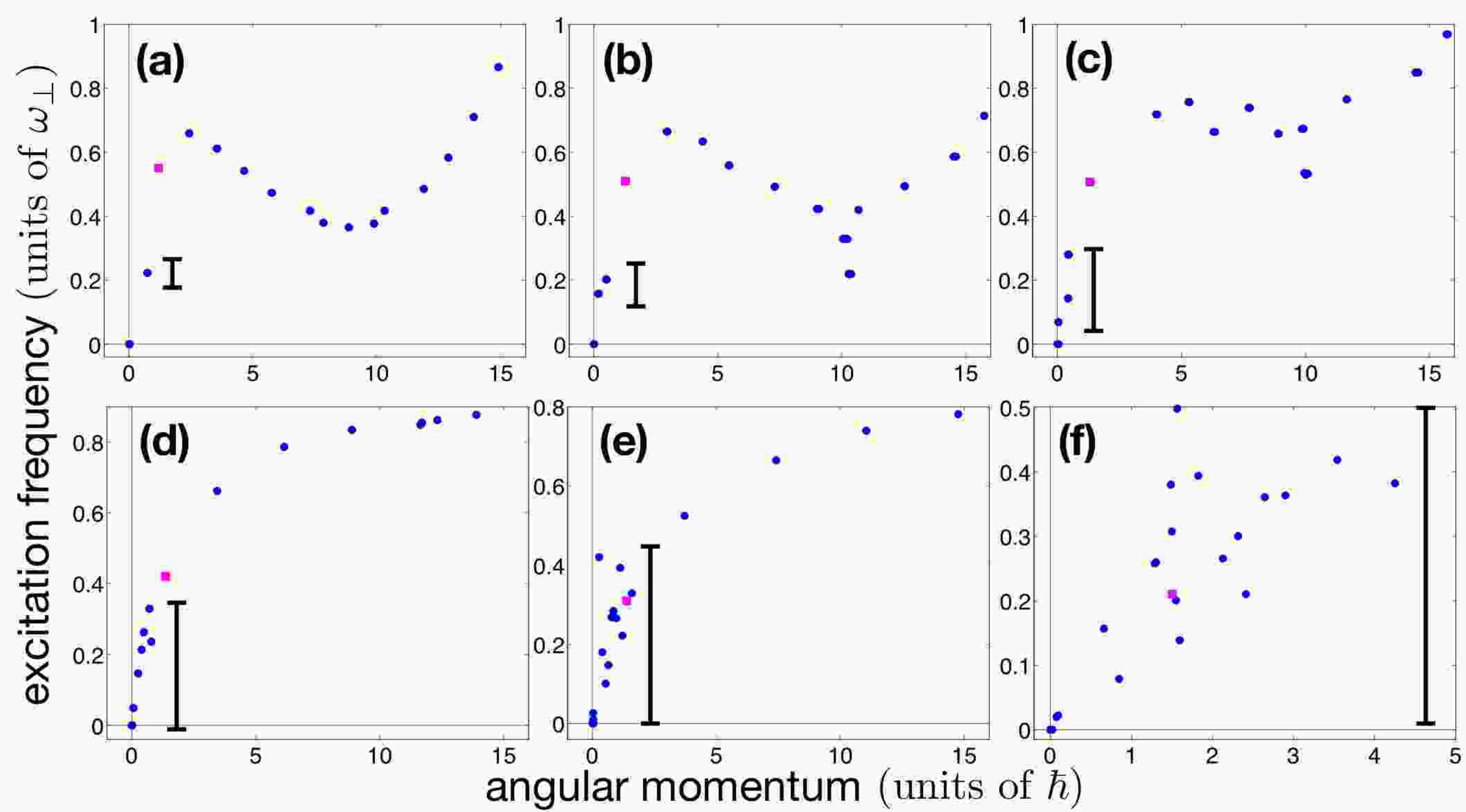}
\caption{(Color online) Calculated low-lying excitation spectra of the respective ground states displayed in Fig.\ref{fig1}. The vertical intervals mark the energy range of the lowest vortex displacement modes. The center-of-mass (Kohn) mode is singled out using a square marker. For every frame there exists a zero mode corresponding to the condensate of atoms. In frame (d) there is an additional zero mode related to the broken continuous rotation symmetry of the ground state.} 
\label{fig2}
\end{figure*}
% FIGURE ===

\section{Excitation spectra}

\begin{figure}
\includegraphics[width=1\columnwidth]{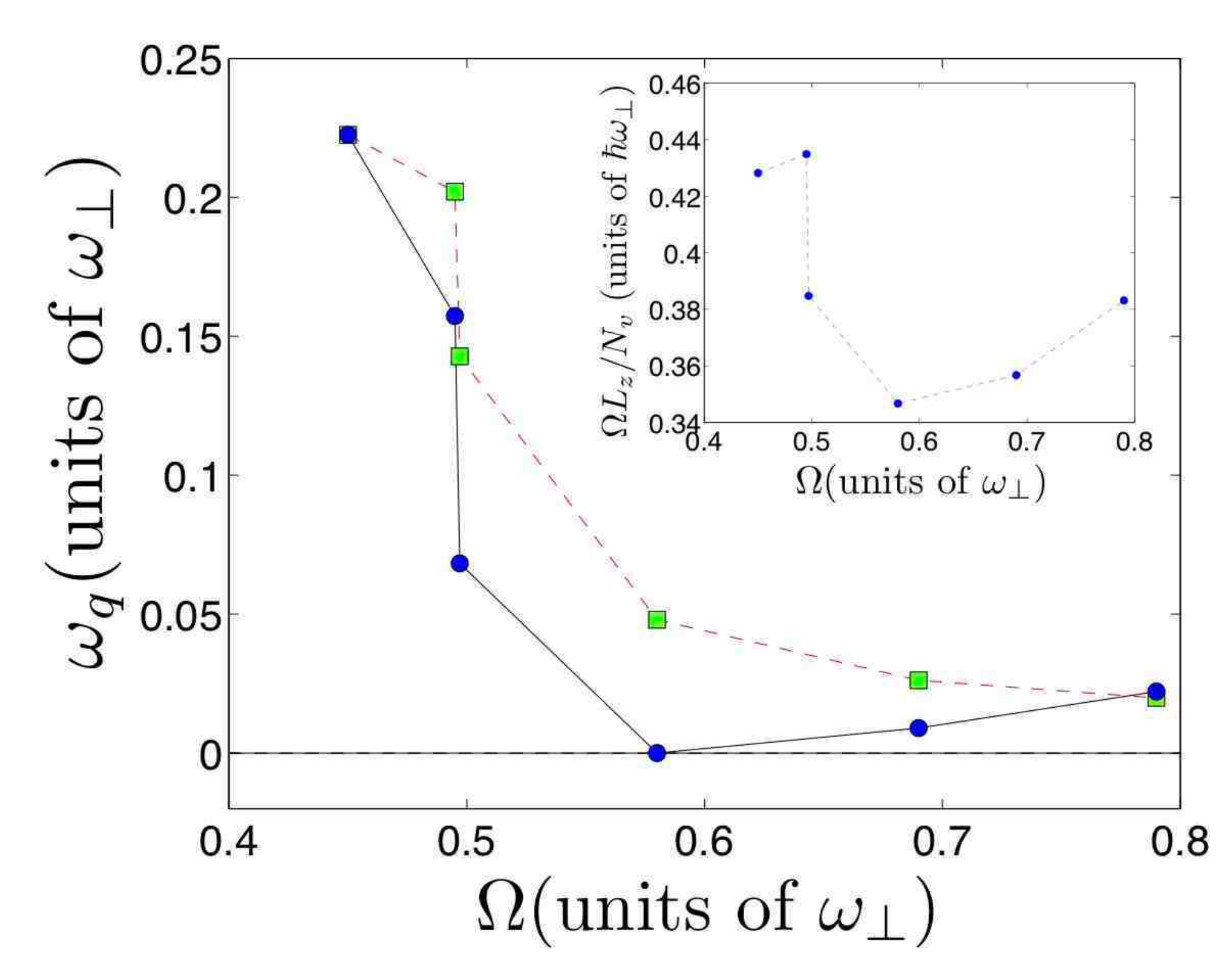}
\caption{(Color online) Excitation frequencies $\omega_q$ of the lowest Tkachenko modes (filled circles) and the lowest common modes (squares) as functions of the external trap rotation frequency $\Omega$. The inset shows the energy per vortex associated with the rotation of the trap as a function of the trap rotation frequency.} 
\label{fig3}
\end{figure}

Figure \ref{fig2} (a)-(f) show the frequencies of the low-lying modes as functions of their orbital angular momentum relative to the condensate ground state, for the systems with $N_v=1,2,3,7,12,$ and $19$ vortices, respectively. The Kelvin--Tkachenko modes listed in Table I are contained within the frequency intervals marked in each frame. The roton-like minima at high angular momenta seed the nucleation of additional vortices to the system when they become resonant with the condensate as the external rotation frequency is increased \cite{Isoshima1999a,Simula2002a,Simula2002b}. These ``hanging" modes are shifted to higher angular momenta as the radial extent of the condensate and the number of vortices in the system increases. The center-of-mass Kohn mode always occurs at frequency $\omega_\perp-\Omega$ and is singled out in the figures using square markers to differentiate from the Kelvin--Tkachenko waves. Each spectrum also contains a zero mode, which corresponds to the Bose--Einstein condensate of atoms. In frame (d) there is an additional zero mode \cite{Simula2012c} shown also in Fig. \ref{fig3} where we have plotted the lowest Tkachenko mode (T) marked with circles and the lowest common modes (C) marked with squares, as functions of the rotation frequency $\Omega$. The frequency of the common mode monotonically decreases with increasing value of $\Omega$. The frequency of the Tkachenko mode first steeply decreases for single orbital vortex states and then increases for multiorbital states. In the limit $\Omega\to \omega_\perp$ the frequency approaches zero \cite{Coddington2003a,Baym2003a}. The inset in Fig. \ref{fig3} shows the rotation energy per vortex $\Omega L_z/N_v$ as a function of the trap rotation frequency $\Omega$. The minimum in this curve is correlated with the excitation frequency of the Tkachenko mode. Such minimum in the dispersion relation for the situations where the system size is comparable to the intervortex spacing has not been predicted by continuum theories, which are valid in the limit of large vortex lattices. The Tkachenko mode with zero oscillation frequency implies this system to possess a doubly degenerate rotating ground state and a finite ground state entropy at $T=0$, unless quantum fluctuations are sufficient to lift this degeneracy. The stability of this zero mode against thermal and quantum fluctuations could be addressed experimentally or by using dynamical finite-temperature methods \cite{Blakie2008a}.

The observed two zero modes suggests the possibility of a co-existence of two Bose--Einstein condensates in the system. The first one results from the spontaneous U(1) gauge symmetry breaking and is the Bose-Einstein condensate of atoms. The second zero energy quasi-particle mode appears due to the continuous SO(2) rotation symmetry breaking. Macroscopic population of such Majorana--Nambu--Goldstone boson would correspond to a Bose--Einstein condensate of Tkachenko waves. Furthermore, for a single vortex system it might be possible to create a condensate of kelvons using suitable trapping parameters. In rotating vector condensates it may be possible to observe further ground state degeneracies such as simultaneous condensates of atoms, magnons and kelvons.

% FIGURE ===
% FIGURE ===

\section{Discussion}

In conclusion, we have calculated the low-lying elementary excitation modes for rotating Bose-Einstein condensates containing up to 19 singly quantized vortex filaments, using the microscopic Bogoliubov--deGennes theory. For $N_v$ vortices in the system, the spectrum contains $N_v$ low-lying Kelvin--Tkachenko collective mode branches. We have created movies of these excitation modes, showing the dynamics of the vortices when these collective modes are populated. In agreement with the classical point vortex model, for every vortex configuration we find two universal modes. These are the fundamental Tkachenko mode and a vortex lattice center-of-mass common mode. For configurations with more than one concentric orbital of vortices, we also find two quadratic modes for which the vortices inside the outermost orbital move together as a rigid body. In addition, several rational modes, for which the central vortex remains stationary, are found. 

These Kelvin--Tkachenko modes could be experimentally observed using the existing experimental cold atom technologies. An interesting future direction is to investigate to what extent the physics of two-dimensional quantum turbulence can be understood in terms of non-adiabatic collective motion of vortices \cite{Virtanen2001d}. For the employed parameters the fundamental Tkachenko mode of the system with seven vortices has a zero oscillation frequency \cite{Simula2012c}. Such zero mode may be interpreted to signal Bose--Einstein condensation of Tkachenko wave quanta in the system. For other studied vortex configurations this Majorana-Nambu-Goldstone mode is found to have finite excitation frequency, lifting the degeneracy observed for the seven-vortex ground state. Further studies of the spectra of slowly rotating Bose-Einstein condensates are needed to investigate the robustness of the observed minimum in the excitation frequency of the lowest Kelvin-Tkachenko mode.

% References ================================================================================

\end{document}